\documentclass{ws-ijmpd}		

\usepackage{amssymb}
\usepackage{latexsym}
\usepackage{amsmath}

\def\ph#1{\phantom{#1}}
\def\hs#1{\hspace*{#1pt}}
\def\ko{\kern1pt}

\begin{document}	

\setcounter{page}{1}

\markboth{V. V. Gurzadyan et al.}
{Ellipticity Analysis of the BOOMERanG CMB Maps}

\catchline{12}{00}{2003}{}{}		

\title{ELLIPTICITY ANALYSIS OF THE BOOMERanG CMB MAPS}

\author{V. G. GURZADYAN,$^{*,\dag}$ P. A. R. ADE,$^{\ddag}$ 
P. De BERNARDIS,$^{\S}$ C. L. BIANCO,$^{\dag}$\\
J. J. BOCK,$^{\ddag}$ A. BOSCALERI,$^{\Vert}$ 
B. P. CRILL,$^{**}$ G. De TROIA,$^{\S}$ K. GANGA,$^{\dag\dag}$\\
M. GIACOMETTI,$^{\S}$ E. HIVON,$^{\dag\dag}$ 
V. V. HRISTOV,$^{**}$ A. L. KASHIN,$^{*}$\\
A. E. LANGE,$^{**}$ S. MASI,$^{\S}$ 
P. D. MAUSKOPF,$^{\ddag}$ T. MONTROY,$^{\ddag\ddag}$\\
P. NATOLI,$^{\S\S}$ C. B. NETTERFIELD,$^{\P\P}$ E. PASCALE,$^{\Vert}$\\
F. PIACENTINI,$^{\S}$ G. POLENTA$^{\S}$ and J. RUHL$^{\ddag\ddag}$}

\address{$^*$Yerevan Physics Institute, Armenia\\
$^\dag$ICRA, Dipartimento di Fisica, Universita' La Sapienza, Roma, Italy\\
$^\ddag$Department of Physics and Astronomy, Cardiff, UK\\
$^\S$Dipartimento di Fisica, Universita' La Sapienza, Roma, Italy\\
$^\P$JPL, Pasadena, USA\\
$^\Vert$IROE-CNR, Firenze, Italy\\
$^{**}$Caltech, Pasadena, USA\\
$^{\dag\dag}$IPAC, Pasadena, USA\\
$^{\ddag\ddag}$Department of Physics, U.C. Santa Barbara, USA\\
$^{\S\S}$Dipartimento di Fisica, Tor Vergata, Roma, Italy\\
and\\
$^{\P\P}$Department of Physics, University of Toronto, Canada}
 
\maketitle


\vspace*{-3pt}

\begin{abstract}
The properties of the Cosmic Microwave Background (CMB) maps carry
valuable cosmo\-logical information. Here we report the results of the
analysis hot and cold CMB anisotropy spots in the BOOMERanG 150~GHz
map in terms of number, area, ellipticity, vs. temperature
threshold. We carried out this analysis for the map obtained by
summing independent measurement channels (signal plus noise map) and
for a compari\-son map (noise only map) obtained by differencing the
same channels. The anisotropy areas (spots) have been identified for
both maps for various temperature thresholds and a \hbox{catalog} of the
spots has been produced. The orientation (obliquity) of the spots is
random for both maps. We computed the mean elongation of spots
obtained from the maps at a given temperature threshold using a simple
estimator. We found that for the sum map there is a region of
temperature thresholds where the average elongation is not dependent
on the threshold. Its value is $\sim2.3$ for cold areas and $\sim2.2$
for hot areas. This is a non-trivial result. The bias of the estimator
is $\lesssim +0.4$ for areas of size $\lesssim 30'$, and smaller for
larger areas. The presence of noise also biases the ellipticity by
$\lesssim +0.3$. These biases have not been subtracted in the results
quoted above. The threshold independent and random obliquity behaviour
in the sum map is stable against pointing reconstruction accuracy and
noise level of the data, thus confirming that these are \hbox{actual}
properties of the dataset.  The data used here give a hint of high
ellipticity for the largest spots. Analogous elongation properties of
CMB anisotropies had been detected for COBE-DMR 4~year data. If this
is due to geodesics mixing, it would point to a non zero curvature of
the Universe.
\end{abstract}


\enlargethispage*{13pt}

\section{Introduction}
The properties of the Cosmic Microwave Background (CMB) radiation
continue to be a key window to our understanding of the early
evolution and the present structure of the Universe. Recent
experiments provided a detection of acoustic peaks in the angular
power spectrum of the CMB, which is an important characteristic of the
conditions at the last scattering epoch, and hence constrains a number
of cosmological parameters (see
e.g. Refs.~\refcite{Bern1,Bern2,Lee2002,DASI2001,CBI,Net} and
\refcite{VSA}). Below we report the results of the analysis of the
ellipticity of the anisotropies in the sky maps from BOOMERanG. At a
given temperature threshold, spots with temperature higher (lower)
than the threshold are identified. There are several definitions of
the ellipticity of one spot, depending on the detailed procedure used
to measure it. Loosely speaking, ellipticity is the ratio between the
major and the minor semi-axes of the ellipse best fitting the contour
of the spot. Ellipticity is expected in the CMB maps, as a result of
the physical effects occurring before recombination, which
\hbox{induce} correlations in the image of the CMB. This paper,
however, is in the spirit of a model independent analysis of the data,
so we will not attempt to compare to and constrain different models of
the CMB anisotropy. Various descriptions have been proposed and
already used for the study of CMB maps. The main aim was\break to
check the Gaussian nature of the data (see
e.g. Refs.~\refcite{Bard86,bond,nov,pol} and
\refcite{pwu}). \hbox{Detailed} numerical simulations of geometrical
descriptions have been carried out in the framework of the Planck
mission in Ref.~\refcite{Barr00}. Particularly, the analysis of the
shapes of anisotropies in the COBE-DMR maps has shown a signature of
elongation,\cite{GT} which was absent in the maps containing only
noise. The present analysis aimed to check the same effect in the
BOOMERanG data. Since these data feature signi\-ficantly higher
angular resolution and lower noise than the COBE-DMR
ones,\cite{ca,torres94a}\break this goal is of particular
interest. 

The elongation analysis, which implies the estima\-tion of
Lyapunov exponents, was initiated with the aim to check the effect of
geodesic mixing,\cite{GK1} which can cause additional elongation of
CMB anisotropies indepen\-dent on the temperature threshold. For the
geodesics in hyperbolic spaces (Anosov systems) the scattering
effectively occurs at every point of the space and in any
two-dimensional directions. The resulting ellipticity has to be
indepen\-dent on the angular size of the spots. This is completely
different from the result of gravitational lensing, the scattering on
clumps of matter, which has been shown to produce a distortion of CMB
spots only at sub-degree scales. The elongation of anisotropies due to
the contribution of the noise, on the other hand, has to be threshold
dependent.\cite{bond} The estimation of Lyapunov exponents,
Kolmogorov--Sinai (KS) entropy and similar geometrical characteristics
has proved its efficiency in the analysis and interpretation of
various experimental results in a variety of fields (see
e.g. Refs.~\refcite{He,So} and \refcite{St}). In the following we show
that the BOOMERanG sum map (A${}+{}$B hereafter, see Fig.~1) reveals a
threshold independent elongation of the anisotropies, which are more
homogeneous with respect to the difference map (A${}-{}$B hereafter,
see Fig.~2). The orientation of the elongations does not show any
preferred direction. The behaviour of ellipticity vs threshold we find
is consistent with the one due to the correlations detected in the
same map by means of the power spectrum analysis.\cite{Net}

\begin{figure}[t]
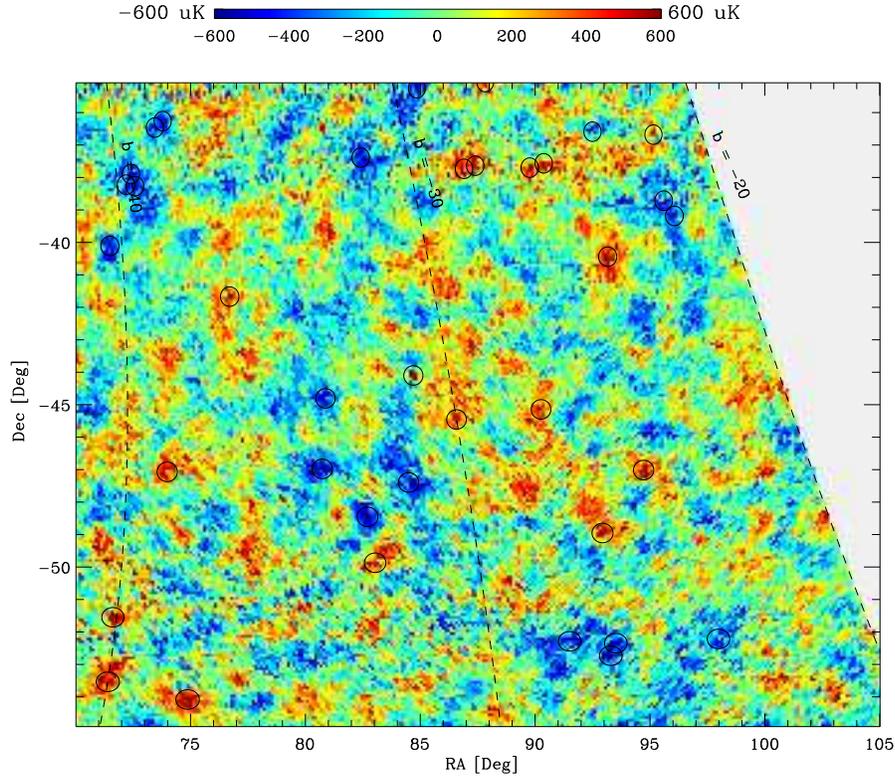
		
\PSFIG{00441f1}{11truecm}{90}
\caption{Sum (A${}+{}$B) map obtained from three independent
measurement channels at 150~GHz: ${\rm A}+{\rm B}= {\rm B150A} + ({\rm
B150A1}+ {\rm B150A2})/2$. The pixel size is 6.9~arc-min (Healpix
nside${}={}$512). The measurement units ($\mu K$) refer to
thermodynamic temperature fluctuations of a 2.73~K Blackbody. The
circles locate the anisotropy spots with more than 3 pixels detected
at a threshold level of $\pm 500~\mu K$ (see Tables~1(a) and 1(b)).}
\end{figure}

\begin{figure}
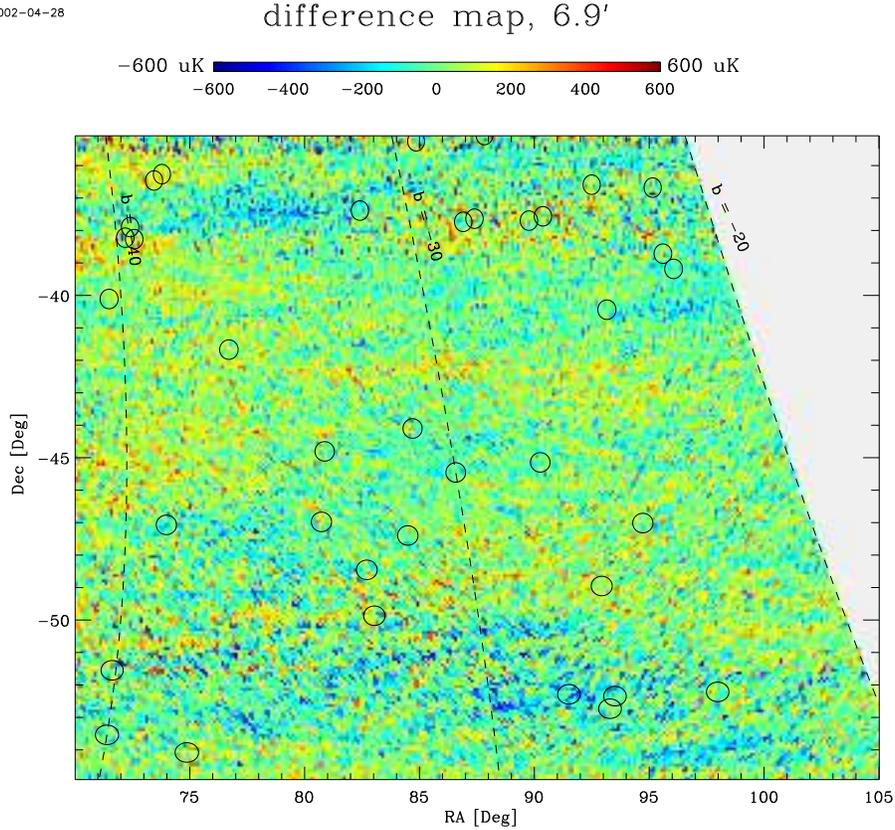
		
\PSFIG{00441f2}{11truecm}{90}
\caption{Difference (A${}-{}$B) map obtained from three independent
measurement channels at 150~GHz: A${}-{}$B =
B150A${}-{}$(B150A1${}+{}$B150A2)/2. The pixel size is 6.9~arc-min
(Healpix nside${}={}$512). The measurement units ($\mu K$) refer to
thermodynamic temperature fluctuations of a 2.73~K blackbody. The
circles locate the anisotropy spots with more than 3 pixels detected
in the sum (A${}+{}$B) map at a threshold level of $\pm 500~\mu K$
(see Tables~1(a) and 1(b)).}
\end{figure}

\section{Complexity of CMB Maps}
The theory of algorithmic information provides tools to study the CMB
maps and extract information on the general properties of the
underlying dynamical systems without specification of cosmological
models. Such a description is the Kolmogorov complexity of the
anisotropy areas (spots)\cite{G} which is the amount of information
required to determine uniquely a given object. The conditional
complexity 
\[
K(x\mid y)= \min[l(p)]
\]
is related to the the amount of information on the object $x$ with
respect the object $y$\cite{Ch},\cite{Kolm} $$
I(y:x)=K(x)- K(x\mid y), $$ and is the minimal length $l(p)$ of the
binary coded program required to describe the object $x$ when the one
for $y$ is known. The complexity is related to Kolmogorov--Sinai (KS)
entropy $h$ via the relation
\begin{equation}	
K_u(t) - K_u(t_0) = \log_2 2^{h(f^t)(t-t_0)}=h(f^t)(t-t_0)\,,
\end{equation}
quantifying the loss of information $\Delta I$ at the evolution of CMB
pattern from the initial state $t_0$ up to $t$, i.e. from the last
scattering epoch up to the observer. The ellipticity of anisotropy
areas at each temperature threshold is the simplest description of the
complexity of the areas. The ellipticity $\epsilon$ is defined via the
divergence of the null geodesics in $(3+1)$-space
\begin{equation}	
\epsilon=\frac{L(t)}{L(t_0)}\,,
\end{equation}
where
\begin{equation}	
L(t) = L(t_0)\frac{a(t)}{a(t_0)} \exp (h s)\,,
\end{equation}
$a(t)$ is the scale factor of the Universe, $s$ is the affine
parameter of the geodesics and KS-entropy is the sum of the positive
Lyapunov exponents;\cite{GK1} for more properties of the geodesics
(see Refs.~\refcite{Anosov,GKbook} and \refcite{LMP}). The geodesic
mixing is a statistical effect arising due to the exponential mixing,
i.e. exponential decay of time correlation functions of the freely
propagating photon beams at $k=-1$, is independent of the conditions
on the last scattering surface, and is distinguished by the threshold
independence and the randomness of the obliquities of the elongated
areas. The ellipticity due to geodesics mixing has to vanish at
precisely flat $k = 0$ and positively curved, $k = +1$,
spaces.\footnote{We avoid the often used terms `open' and `closed'
Universe, since the geometry does not define the topology, e.g. the
flat, $k = 0$ Universe can be not only $R^3$ but also $S^1 \times
R^2$, $Tor^3$, $R^1 \times Tor^2$, etc., and similarly the negatively
curved $k = -1$ Universe can be both `open' and `closed'.}

\section{Data}
BOOMERanG is a millimetric telescope with bolometric detectors on a
balloon borne platform.\cite{Crill02,Pia} It was flown in 1998/99 and
produced wide (4$\%$ of the sky), high resolution ($\sim 10'$) maps of
the microwave sky (90 to 410~GHz).\cite{Bern1} Two observa\-tion modes
were used to map this sky patch: sky scans at a speed of 1$^\circ$/s
and sky scans at 2$^\circ$/s. This allows us to perform powerful tests
for systematics.\cite{Net} In fact, at 2$^\circ$/s, the sky
temperature distribution produces signals in the detection chain at
\hbox{frequencies} which are doubled with respect to observations at
1$^\circ$/s, while instrument related effects, like 1/f noise,
microphonic lines, and time-domain \hbox{response} remain at the same
frequency. Comparing the maps obtained in the two observation modes is
thus very effective in detecting instrumental artifacts. In the
BOOMERanG maps the CMB structure is resolved with high signal to noise
ratio, and hundreds of degree-scale hot and cold areas are
evident. The rms temperature fluctuation of these areas is $\sim
80~\mu K$. The detected fluctuations are spectrally consistent with
the derivative of a 2.735~K blackbody.\cite{Masi3} have shown that
contamination from \hbox{local} foregrounds is negligible in the maps
at 90, 150 and 240~GHz, and that the 410~GHz channel is a good monitor
for dust emission. In our study we used two maps at 150~GHz. These
maps have been obtained from the time ordered data using an iterative
procedure,\cite{Nato01} which properly takes into account the system
noise and produces a maximum likelihood map. The largest structures
(scales larger than 10$^\circ$) are removed in this procedure, to
avoid the dominating effects of instrument drifts and $1/f$ noise. The
two input maps, A and B, included 33111 pixels, each of $\sim
7$~arc-min in linear size, in a high Galactic latitude region covering
about 1\% of the sky, with coordinates $RA > 70^{\circ}$, $-55^{\circ}
< {\rm dec} < -35^{\circ}$ and $b < -20^\circ$.\cite{Bern1} The first
map (A) has been obtained from the data of the B150A detector, while
the second map (B) was obtained by averaging the maps from detectors
B150A1 and B150A2. In this way we obtained two maps with similar noise
per pixel. The (CMB) signal to noise ratio per pixel is of the order
of 1 for our $7'$ pixels. The sum (A${}+{}$B) and difference
(A${}-{}$B) maps from all scans (1$^\circ$/s and 2$^\circ$/s) are
shown in Figs.~1 and 2 respectively. There are three AGN with
significant flux in the maps (double circles in Figs.~1 and 2). This
has been taken into account in the analysis.

\vspace*{-1pt}

\section{Analysis}
We studied the excursion sets in the BOOMERanG maps by means of a
specially developed software.\cite{Kashin} This enabled, in an
interactive way, to change the input parameters, like the threshold
level and the minimum and maximum number of pixels forming an
anisotropy area to be included in the analysis, and allowed the
visualization of all the intermediate steps of the analysis.

\enlargethispage*{13pt}

\vspace*{-1pt}

\subsection{Algorithm}
To define the excursion sets (hereafter `areas') in the maps, a matrix
of temperature data of the pixels with equal and higher than the given
temperature threshold (lower, for negative thresholds) has been formed
and the contours of those areas have been studied. The original maps
follow the Healpix pixelization scheme.\cite{Gors1998} However, the
distance between pixel centers is not constant for this pixelization,
so we had to regrid the maps. We also oversampled the maps in order to
easy the algorithm of definition of the excursion sets. We have
reprojected the data using both Cartesian and curvilinear coordinates.
We found that for our purpose the two are equivalent, and we used the
Cartesian coordinates for simplicity. This procedure enabled us to
check the sensitivity of the ellipticity results with respect to the
coordinate system and the cell size, for different total numbers of
cells. In particular, an oversampled 1~arc-min cell (a $1692\times
1296$ matrix) appears to lead to more accurate results than the
original 7.5~arc-min cell matrix. The procedure of the definition of
the centers of areas, their semi-axes and their {\it obliquity} and
{\it ellipticity} was as follows:
\begin{arabiclist}[(1)]
\item[(1)] Center of the area. We define the coordinates of the center
as $y_c=(y_2-y_1)/2$, $x_c=(x_2-x_1)/2$, where $y_1$ and $y_2$ are the
highest and lowest $y$ coordinates of the pixels of the area (and
analogously for $x$ coordinates).

\item[(2)] Semi-major axis $d_{\max}$. It is estimated as the segment
connecting $(x_c, y_c)$ and the center of the farthest pixel of the
area. The inclination angle of the segment is the {\it obliquity}, and
is measured counter-clock-wise from the positive $x$ semi-axis
(parallel to RA). 

\item[(3)] Semi-minor axis $d_{min}$. On either sides of the major
axis we find the pixels having the maximum distance between the pixel
center and the major axis. We take the average of the two distances as
the length of the semi-minor axis.

\item[(4)] The {\it ellipticity} is computed as $\epsilon=
d_{\max}/d_{\min}$.
\end{arabiclist}

For each temperature threshold the mean ellipticity of the anisotropy
areas was estimated for both the A${}+{}$B and A${}-{}$B maps, along
with the angular distribution of the obliquities. Our algorithm to
estimate ellipticity is simpler and faster than the one used in
Ref.~\refcite{Barr00}, but, in the presence of noise, is biased, as we
show below. The level of biasing, however, is reasonably small, and
completely acceptable for our purpose.

\begin{figure}[b]
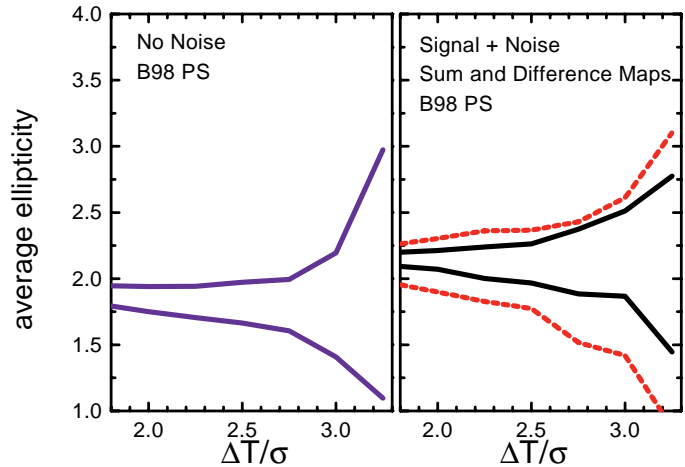
		
\epsfxsize=9truecm
\figurebox{}{}{00441f3}
\caption{Average ellipticity of intrinsic CMB anisotropy (left)
obtained from simulations of CMB maps with the best fit power spectrum
measured by BOOMERanG. The two lines define the 68\% confidence
interval for the map-averaged ellipticity measured from the
simulations. In the right part of the figure we add realistic noise
and filtering, at the same level present in the BOOMERanG 150~GHz
channels, and analyze the map-average ellipticity of simulations
obtained summing (continuous lines) and differencing (dashed lines)
maps from two independent measurement channels. The presence of noise
increases the average ellipticity in the maps by $\sim 0.3$. In the
difference maps, where only noise is present, the scatter of
map-averaged ellipticities is higher.}
\end{figure}

\subsection{Simulations}
Since the use of ellipticity is relatively new in the CMB literature,
we carried out numerical simulations in order to show the performance
of our estimator and the expected behaviour for the map of the CMB. In
order to validate our algorithm we produced simulated maps, with
circular, symmetric Gaussians, well separated in the sky, with FWHM
$\sim30$~arc-min. We used the same pixelization used for the BOOMERanG
data. The average ellipticity we measure is between 1.3 (lower
thresholds) and 1.4 (higher thresholds). The deviation from the
expected unit value is due to pixelization and to the algorithm used
to define ellipticity. The result gets closer to 1 for larger FWHM. If
we use Gaussians with two axis (minor 30~arc-min, major 45~arc-min),
and we keep them well separated, we get an ellipticity of $\sim 1.8$,
probably due to the bias we have seen before. Finally, we moved
randomly the symmetric Gaussians, so that some of them merged and
produced elliptical spots. In this case the mean ellipticity is a
function of threshold, as expected. We conclude that our estimator of
ellipticity is biased at a level of $\lesssim +0.4$ for spots smaller
than $30'$, and $\lesssim +0.1$ for spots larger than 1$^\circ$. In
Fig.~3 we illustrate the expected ellipticity behaviour of intrinsic
CMB ellipticity in the context of the currently \hbox{popular} adiabatic
inflationary model. Many realization of CMB maps of the sky region
\hbox{observed} by BOOMERanG have been simulated, starting from the best fit
angular power spectrum measured by BOOMERanG. For different
temperature thresholds $T_i$, the anisotropy spots hotter than $T_i$
have been identified, and their ellipticities have been computed. In
the left part of Fig.~3 the two lines define the 68\% confidence
intervals for the map-averaged ellipticity derived from these
simula\-tions. A map-averaged ellipticity $\sim 2$ is expected,
basically independent of the temperature threshold. In the right part
of Fig.~3 we simulate the presence of instrumental noise in the
measurements of ellipticity, and show how to extract the CMB
ellipticity signal from noisy sky maps. We have added realistic noise
and filtering, at the same level present in the BOOMERanG 150~GHz
channels, and analyzed the map-average ellipticity of simulations
obtained summing and differencing maps from two independent
measurement channels. The presence of noise increases the level of the
average ellipticity in the sum maps, at a level of about
0.3. In the difference maps, where only noise is present, the scatter
of map-averaged ellipticity is higher, as expected.\cite{bond}

\renewcommand{\thetable}{\arabic{table}(\alph{enumii})}	
\setcounter{table}{0}	
\setcounter{enumii}{1}

\begin{table}[b]		
\tbl{Threshold: $-500~\mu K$, A + B.} 
{\begin{tabular}{@{}ccccc@{}}
\hline\\[-7pt]
Area & \multicolumn{2}{c}{Coordinates, degree} & Number & Ellipticity \\[3pt]
No & \hs{12}$\ell$ & $b$ & of pixels & \\[3pt]
\hline\\[-7pt]
C1\ph{0} & \hs{8}243.9 & $-40.4$ & 5 & 4.72\\ 
C2\ph{0} & \hs{8}241.5 & $-39.7$ & 3 & 1.42\\ 
C3\ph{0} & \hs{8}241.1 & $-39.5$ & 4 & 1.96\\ 
C4\ph{0} & \hs{8}241.6 & $-39.4$ & 3 & 2.43\\ 
C5\ph{0} & \hs{8}239.4 & $-38.5$ & 3 & 1.61\\ 
C6\ph{0} & \hs{8}239.2 & $-38.2$ & 4 & 4.33\\ 
C7\ph{0} & \hs{8}253.1 & $-34.2$ & 12 & 1.73\\
C8\ph{0} & \hs{8}250.5 & $-33.9$ & 6 & 2.20\\ 
C9\ph{0} & \hs{8}255.0 & $-33.0$ & 12 & 2.21\\
C10 & \hs{8}253.9 & $-31.7$ & 8 & 1.69\\ 
C11 & \hs{8}242.0 & $-31.6$ & 6 & 2.49\\ 
C12 & \hs{8}240.1 & $-29.2$ & 4 & 3.17\\ 
C13 & \hs{8}260.2 & $-27.9$ & 6 & 2.24\\ 
C14 & \hs{8}260.9 & $-26.9$ & 4 & 2.39\\ 
C15 & \hs{8}260.5 & $-26.7$ & 7 & 2.02\\ 
C16 & \hs{8}261.0 & $-24.0$ & 3 & 1.28\\ 
C17 & \hs{8}243.4 & $-23.6$ & 4 & 2.18\\ 
C18 & \hs{8}246.4 & $-21.9$ & 3 & 2.48\\ 
C19 & \hs{8}247.0 & $-21.7$ & 4 & 1.95\\[3pt] 
\hline
\end{tabular}}
\end{table}

\setcounter{table}{0}	
\setcounter{enumii}{2}

\begin{table}		
\tbl{Threshold: $+500~\mu K$,  A + B.}
{\begin{tabular}{@{}ccccc@{}}
\hline\\[-7pt]
Area & \multicolumn{2}{c}{Coordinates, degree} & Number & Ellipticity \\
No & \hs{12}$\ell$ & $b$ & of pixels & \\[3pt]
\hline\\[-7pt]
 H1\ph{0} & \hs{8}258.9 & $-40.1$ & 5 & 2.10\\ 
 H2\ph{0} & \hs{8}261.5 & $-40.0$ & 3 & 1.60\\ 
 H3\ph{0} & \hs{8}253.0 & $-38.8$ & 4 & 1.86\\ 
 H4\ph{0} & \hs{8}261.9 & $-37.9$ & 5 & 2.52\\ 
 H5\ph{0} & \hs{8}246.3 & $-36.6$ & 4 & 2.25\\ 
 H6\ph{0} & \hs{8}256.7 & $-32.9$ & 3 & 2.84\\ 
 H7\ph{0} & \hs{8}240.7 & $-32.7$ & 3 & 2.39\\ 
 H8\ph{0} & \hs{8}250.1 & $-31.1$ & 5 & 1.68\\ 
 H9\ph{0} & \hs{8}251.9 & $-30.0$ & 3 & 1.45\\ 
 H10 & \hs{8}243.3 & $-28.2$ & 5 & 1.89\\ 
 H11 & \hs{8}243.3 & $-27.8$ & 4 & 2.39\\ 
 H12 & \hs{8}252.1 & $-27.4$ & 3 & 1.57\\ 
 H13 & \hs{8}240.6 & $-26.8$ & 3 & 1.53\\ 
 H14 & \hs{8}256.7 & $-26.4$ & 4 & 1.71\\ 
 H15 & \hs{8}243.9 & $-26.0$ & 6 & 2.27\\ 
 H16 & \hs{8}243.9 & $-25.5$ & 3 & 2.41\\ 
 H17 & \hs{8}254.9 & $-24.8$ & 6 & 1.62\\ 
 H18 & \hs{8}247.6 & $-24.2$ & 7 & 1.61\\ 
H19 & \hs{8}244.2 & $-21.6$ & 3 & 2.16\\[3pt] 
\hline
\end{tabular}}
{\bf Note to Table~1(b):} Area H7 includes a known AGN. Other two
AGNs are present in the map we have analyzed, but they fill less
than 3 pixels and were discarded.
\end{table}

\begin{figure}
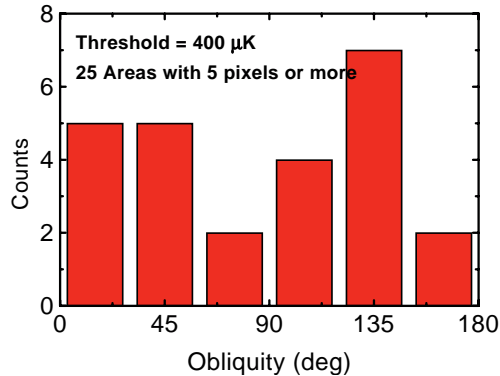
		
\vspace*{10pt}
\epsfxsize=6.5truecm
\figurebox{}{}{00441f4}
\caption{Typical obliquity histogram for the A${}+{}$B map. The
threshold is $+400~\mu K$, and only areas with more than 4 pixels
have been included.}
\end{figure}

\setcounter{table}{1}
\setcounter{enumii}{1}

\subsection{Application to the BOOMERanG data}
Our software enabled to explicitly follow the evolution of anisotropy
areas with respect the temperature threshold and the role of each area
in the final results. As an example, Table~1 contains the data for the
hot and cold anisotropy areas at thresholds $\pm 500~\mu K$.  In
Fig.~4 we plot a histogram of obliquity of the spots detected in the
sum map A${}+{}$B, at a temperature threshold of $+400~\mu K$. Only
areas formed by 5 pixels or more have been considered, because the
obliquity error due to pixelization is very large for areas with less
pixels. The obliquities of these 25 areas are random. The $\chi^2$ for
a uniform distribution is 6.13 with 5~DOF. The same typical behaviour
has been found at different thresholds.  Only when areas with
relatively small \hbox{number} of pixels are included, there is a certain
domination of alignment on 45 and 135 degrees, which disappears when
such areas are abandoned, thus indicating the role of the rectangular
shapes of the pixels themselves. This test is a strong indication that
the origin for the detected ellipticity cannot be
instrumental. Effects related to detectors time constants or to the
scan strategy should be strongly anisotropic, as our scans are all
within $\pm 12^\circ$ from the DEC = constant lines.  In Fig.~5 we
plot the map-averaged ellipticity versus the temperature threshold for
both the A${}+{}$B map (signal plus noise) and the A${}-{}$B map
(noise only). In Fig.~6 we plot the same for a previous data release
with higher noise ($S/N$ of the order of 0.7 versus 1.0) and lower
accuracy pointing reconstruction ($4.5'$~rms versus $2.5'$~rms). These
data have been considered in order to investigate if the results are
robust against variations of the noise level and of the pointing
accuracy. The use of perturbed data is a standard and powerful
technique in the framework of theory of dynamical
systems.\cite{Arnold} We use it in order to check at once the effect
of perturbations like inaccuracy in pointing, timestream filtering,
detector noise, which are different for the two data releases we
compare. Only the areas containing 3 to 200 pixels have been used,
since areas with 1 and 2 pixels introduce biases related to the shape
of the pixels. The mean elongation has been computed only for
thresholds containing at least two areas. Including areas with more
than 200 pixels is not informative, since only a few of such
structures are present, and their topological properties are expected
to be different from those of smaller areas. The lower boundary (in
absolute value) of the temperature threshold interval of interest was
determined as the level where most of the areas already have shapes
for which an ellipticity can be assigned. The upper boundary marks the
threshold where the number of areas present in the map is not less
than 3. In Table~2 more detailed information on the properties of the
areas as a function of the temperature threshold is given. In addition
to ellipticity we report the number of areas versus threshold and the
average size of the areas. The existence of a threshold independent
region for the ellipticity from the A${}+{}$B map is seen in both
Fig.~5 (referring to the final maps) and Fig.~6 (referring to the
higher noise and less accurate pointing). This means that this feature
is robust against \hbox{perturbations} of the pointing and noise level. The
data with higher noise level enabled one to compare some properties of
A${}+{}$B and A${}-{}$B, to reveal the role of the noise. We proceeded
as follows. We defined `equivalent' sets of areas for the A${}+{}$B
and A${}-{}$B maps, respectively, by selecting the two temperature
threshold intervals with the same number of areas contributing to our
measurement of ellipticity. For example, for positive thresholds, the
interval $[450,625]$ in the A${}+{}$B map must be compared to the
interval $[300,400]$ in the A${}-{}$B map, since both intervals
include from 5 to over 50 areas. This scheme defines threshold
intervals $[-300, -450]$~$\mu K$ and $[300, 400]$~$\mu K$ for
A${}-{}$B and $[-475, -625]$~$\mu K$ and $[450, 625]$~$\mu K$ for
A${}+{}$B, as seen in Fig.~5. We compute the variance of the
ellipticity in those threshold intervals, as
$\sigma^2=\sum(\epsilon-\bar{\epsilon})^2/(N-1)$, where $N$ is the
number of thresholds in the interval. We find that the peak to peak
scatter, for negative thresholds, is over $5\sigma$ for A${}-{}$B,
while is $3\sigma$ for A${}+{}$B, with the overwhelming majority of
points inside the $2\sigma$ interval. Another feature not evident from
Figure~5 is the scatter in ellipticities at a given threshold. In the
first case (A${}+{}$B) the distribution of areas ellipticities for a
given threshold is in general narrower than in the second case
(A${}-{}$B) especially at high thresholds. For example,
$\delta\epsilon = \epsilon_{\max}-\epsilon_{\min}$ equals 1.3 for the
threshold corresponding to 6 areas in A${}+{}$B, (-600 $\mu K$), while
it is 2.8 for the threshold corresponding to 5 areas in A${}-{}$B
(-400 $\mu K$). In other words, at high thresholds the areas in
A${}+{}$B possess more homogeneous ellipticities with respect to the
areas in A${}-{}$B. It is remarkable that in their other properties
the A${}+{}$B and A${}-{}$B areas show no qualitative
difference. Namely, they show very similar properties in: (A) the
variation of the number of spots vs threshold, (B) the dependence of
the mean area sizes, in pixel numbers per area, on the threshold
(which is very similar within the ``equivalent" intervals), (C) the
ellipticity dependence on the area size. It thus appears that the two
sets have identical properties except for the scatter of ellipticity
over the thresholds. This behavior of ellipticity is robust against
the variation of the parameters. For example, the change of the
minimal number of the pixels per area from 3 to 5, 7, 10 does not
change the properties of elongation in the A${}+{}$B data. We also
selected only the spots larger than 100 pixels (corresponding to few
degrees sized spots) and found again an ellipticity $2.3 - 2.5$ with
the same stability with respect to threshold we found for smaller
spots. For these spots the bias due to noise and algorithm is smaller
than $+0.3$. The grand-average ellipticities for the final maps are
reported in Table~3.

\begin{figure}[t]
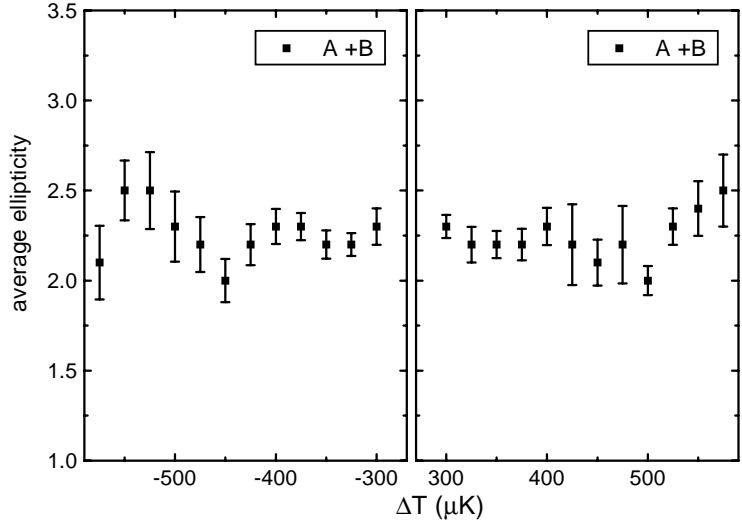
		
\epsfxsize=10truecm
\figurebox{}{}{00441f5}
\caption{Ellipticity vs temperature threshold (in $\mu K$) for sum
(A${}+{}$B) map at 150~GHz. Only areas containing more than 3 pixels
and less than 200 in the final maps have been considered.}
\end{figure}
\begin{figure}
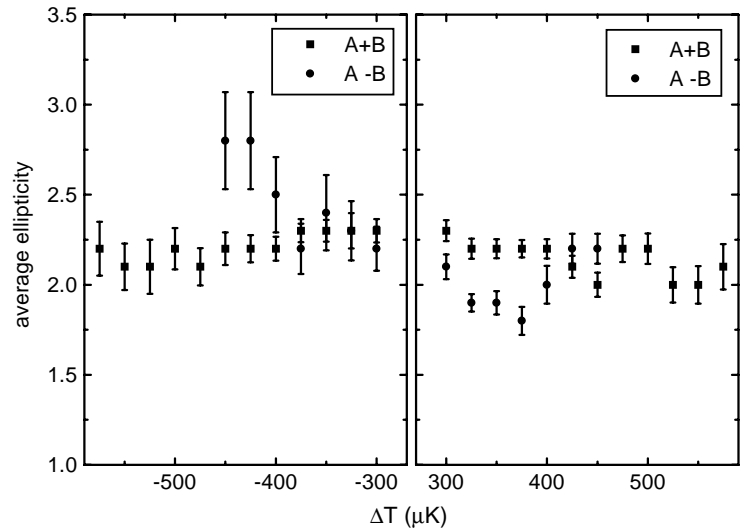
		
\epsfxsize=10truecm
\figurebox{}{}{00441f6}
\caption{Same as Fig.~5 for data with preliminary pointing
reconstruction ($4.5'$~rms pointing error instead of $2.5'$) and lower
signal to noise ratio (0.7 instead of 1). Data for the difference maps
are also plotted as circles.}
\end{figure}
\begin{table}[b]	
\tbl{Pixel range: 3--200, A + B thresholds $< 0$.}
{\begin{tabular}{@{}ccccccc@{}}
\hline\\[-7pt]
Threshold $(\mu K)$ & Pixels & Areas & Pix/Ar & $\epsilon_{\rm min}$ 
& $\epsilon_{\rm max}$ & $\epsilon_{\rm mean}$ \\[3pt]
\hline\\[-7pt]
$-300$ & 1669 & 180 & 9.30 & 1.28 & 6.72 & 2.25\\
$-325$ & 1236 & 152 & 8.10 & 1.25 & 4.38 & 2.16\\
$-350$ & 947\ph{0} & 125 & 7.60 & 1.25 & 4.83 & 2.18\\
$-375$ & 727\ph{0} & 105 & 6.90 & 1.28 & 4.37 & 2.26\\
$-400$ & 538\ph{0} & 81\ph{0} & 6.60 & 1.28 & 4.82 & 2.27\\
$-425$ & 364\ph{0} & 59\ph{0} & 6.20 & 1.28 & 4.82 & 2.19\\
$-450$ & 262\ph{0} & 44\ph{0} & 6.00 & 1.28 & 4.54 & 2.05\\
$-475$ & 170\ph{0} & 31\ph{0} & 5.50 & 1.28 & 4.72 & 2.24\\
$-500$ & 101\ph{0} & 19\ph{0} & 5.30 & 1.28 & 4.72 & 2.24\\
$-525$ & 73\ph{00} & 15\ph{0} & 4.90 & 1.42 & 4.72 & 2.54\\
$-550$ & 39\ph{00} & 10\ph{0} & 3.90 & 1.28 & 3.40 & 2.46\\
$-575$ & 22\ph{00} & 6\ph{00} & 3.70 & 1.42 & 3.40 & 2.14\\
$-600$ & 9\ph{000} & 3\ph{00} & 3.00 & 1.56 & 2.48 & 2.08\\
$-625$ & 3\ph{000} & 1\ph{00} & 3.00 & 2.19 & 2.19 & 2.19\\
$-650$ & 3\ph{000} & 1\ph{00} & 3.00 & 2.19 & 2.19 & 2.19\\[3pt]
\hline
\end{tabular}}
\end{table}

\setcounter{table}{1}
\setcounter{enumii}{2}
\begin{table}		
\tbl{Pixel range pixel: 3--200, A + B thresholds $> 0$.} 
{\begin{tabular}{@{}ccccccc@{}}
\hline\\[-7pt]
Threshold $(\mu K)$ & Pixels & Areas & Pix/Ar & $\epsilon_{\rm min}$ 
&  $\epsilon_{\rm max}$ & $\epsilon_{\rm mean}$ \\[3pt]
\hline\\[-7pt]
300 & 1601 & 187 & 8.60 & 1.15 & 4.60 & 2.26\\
325 & 1196 & 153 & 7.80 & 1.15 & 5.97 & 2.18\\
350 & 899\ph{0} & 134 & 6.70 & 1.29 & 4.77 & 2.20\\
375 & 623\ph{0} & 100 & 6.20 & 1.29 & 4.82 & 2.17\\
400 & 426\ph{0} & 76\ph{0} & 5.60 & 1.29 & 4.93 & 2.30\\
425 & 286\ph{0} & 51\ph{0} & 5.60 & 1.28 & 7.66 & 2.23\\
450 & 178\ph{0} & 37\ph{0} & 4.80 & 1.28 & 4.37 & 2.09\\
475 & 119\ph{0} & 25\ph{0} & 4.80 & 1.34 & 5.57 & 2.16\\
500 & 79\ph{00} & 19\ph{0} & 4.20 & 1.45 & 2.84 & 1.99\\
525 & 46\ph{00} & 12\ph{0} & 3.80 & 1.45 & 2.84 & 2.25\\
550 & 25\ph{00} & 7\ph{00} & 3.60 & 1.60 & 3.27 & 2.44\\
575 & 15\ph{00} & 4\ph{00} & 3.80 & 1.60 & 3.21 & 2.51\\
600 & 6\ph{000} & 2\ph{00} & 3.00 & 1.42 & 1.60 & 1.51\\
625 & 3\ph{000} & 1\ph{00} & 3.00 & 1.42 & 1.42 & 1.42\\
650 & 3\ph{000} & 1\ph{00} & 3.00 & 1.42 & 1.42 & 1.42\\[3pt]
\hline
\end{tabular}}
\end{table}

\renewcommand{\thetable}{\arabic{table}}	
\setcounter{table}{2}	

\begin{table}[t]		
\tbl{}
{\tabcolsep12pt\begin{tabular}{@{}lcc@{}}
\hline\\[-7pt]
Pixel interval per area & $[3,200]$ & $[3,200]$ \\
Threshold interval ($\mu K$) & $[-400,-600]$ & $[375,575]$ \\
Number of areas & 268 & 331 \\
Mean ellipticity & $2.32\pm 0.06$ & $2.22\pm 0.05$ \\[3pt]
\hline
\end{tabular}}
{\bf Note to Table~3:} The quoted errors are statistical only. A
bias of $\lesssim 0.3$ due to the algorithm, and a similiar bias due
to the noise have not been subtracted.
\end{table}

\subsection{Systematics}
Possible systematic effect could be expected for very small areas due
to the shape of the Healpix pixels. However, we find that this effect,
for such small spots, is smeared by the intrinsic inaccuracy of the
definition of the two semi-axes.  We also checked the effect of
including data closer to the Galactic plane (down to
$b<-13^\circ$). We found that the flat ellipticity vs threshold
behaviour was strongly distorted by the presence of low galactic
latitude data.  We repeated the analysis above separately for maps
obtained from 1$^\circ$/s scans only and for maps obtained from
2$^\circ$/s only, and obtained consistent results. The difference map
(1$^\circ$/s map --- 2$^\circ$/s map) features a larger scatter of the
map-averaged ellipticity vs threshold diagram, as expected for noise
only. This test strongly excludes an instrumental origin of the
ellipticity of the areas.

\section{Conclusions}
We have produced a catalog of hot and cold spots in the BOOMERanG CMB
maps. The distribution of number of spots and of their areas vs
temperature threshold has been computed and can be compared to the
predictions of cosmological models. We also found a threshold
independent elongation of the spots in the meaningful interval of
temperature thresholds. Selecting areas with three and more pixels,
within temperature thresholds for $[375, 575]$ $\mu K$ (hot areas), we
obtained a mean ellipticity $2.22 \pm 0.05$; in the range
$[-400,-600]$ $\mu K$ (cold areas) we obtained a mean ellipticity
$2.32 \pm 0.06$. The quoted errors are statistical only.  

This is the
first measurement of the average ellipticity of the CMB at this
angular resolution. The fact that its value is around 2 and not 1 or 3
should be regarded as a non-trivial experimental result.  The bias
deriving from pixelization and algorithm is of the order of $+0.3$,
while the bias due to the noise is also $\sim 0.3$ for areas smaller
than $30'$. The ellipticities quoted above have not been corrected for
these effects. For larger spots ($\sim 2^\circ$ size) the bias is
smaller but the ellipticity remains $\sim 2.5$.  This means that for
large areas the measured ellipticity is slightly higher than expected
in the standard model.  

The obliquities of anisotropies are
random. The threshold independent and random obliquity behaviors in
the sum map are stable against pointing reconstruction accuracy and
noise level of the data, thus confirming that these are actual
properties of the dataset. The same description estimated for the COBE
maps yielded $\epsilon_{\rm COBE} = 1.88 \pm 0.17$.\cite{GT} The
analysis described here, however, was based on a much larger number of
areas than in COBE, and an adaptive software was used.  

A priori there is no reason for a random field to produce a constant
ellipticity vs threshold. The detected behavior has to be produced by
the correlations present in the sky signal. The presence of
correlations in the map is not surprising, as evident from the power
spectrum of the maps,\cite{Net} and indeed our experimental results
are consistent with the results of simulations (compare Fig.~3 to
Fig.~5). 

There is significant debate about the nature of the
cosmological model after the recent WMAP
release.\cite{Aur03,Bri03,Ben03,Efs03,Luminet,Ell03} Non-precisely
zero curvature is among the discussed reasons of the low power at low
multipoles of the CMB. The accuracy of the data used here gives also a
hint of high ellipticity for the largest spots. If this is due to
geodesics mixing, it would also point to a non-zero curvature.  This
analysis can be within reach of the data from forthcoming experiments.

\section*{Acknowledgments}
The BOOMERanG experiment is supported in Italy by Agenzia Spaziale
Italiana, Programma Nazionale Ricerche in Antartide, Universita' di
Roma La Sapienza; by PPARC in the UK, by NASA, NSF OPP and NERSC in
the U.S.A., and by CIAR and NSERC in Canada.

\def\x#1#2{#2 #1}

\end{document}